\documentclass[12pt]{article}
\usepackage{amsmath,amsfonts,amssymb,latexsym}
\usepackage{epsfig}
\usepackage{graphicx}
\usepackage{wrapfig}
\usepackage{subfigure}
\usepackage{cite}

\newcommand{\sect}[1]{\setcounter{equation}{0}\section{#1}}
\newcommand{\subsect}[1]{\subsection{#1}}

\def\ds{\displaystyle}
\def\be{\begin{equation}}
\def\ee{\end{equation}}
\def\bea{\begin{eqnarray}}
\def\eea{\end{eqnarray}}
\def\bean{\begin{eqnarray*}}
\def\eean{\end{eqnarray*}}
\def\N{{\mathbb N}}
\def\Z{{\mathbb Z}}
\def\R{{\mathbb R}}

\def\E{{\mathbb E}}
\def\a{{\alpha}}
\def\b{\beta}
\def\G{\Gamma}

\begin{document}


\thispagestyle{empty}
\hfill \today

\vspace{2.5cm}

\begin{center}
\bf{\LARGE 
Jacobi polynomials and $SU(2,2)$ 
}
\end{center}

\bigskip\bigskip

\begin{center}
E. Celeghini$^1$, M.A. del Olmo$^2$ and M.A. Velasco$^2$\footnote{Present address: {\sl CIEMAT, 
Madrid, Spain.}}
\end{center}

\begin{center}
$^1${\sl Dipartimento di Fisica, Universit\`a  di Firenze and
INFN--Sezione di
Firenze \\
I50019 Sesto Fiorentino,  Firenze, Italy}\\
\medskip

$^2${\sl Departamento de F\'{\i}sica Te\'orica and IMUVA, Universidad de
Valladolid, \\ 
E-47011, Valladolid, Spain.}\\
\medskip

{e-mail: celeghini@fi.infn.it, olmo@fta.uva.es}

\end{center}

\bigskip

\bigskip

\begin{abstract}
A ladder structure of operators is presented for the Jacobi polynomials, $J_{n}^{(\a,\b)}(x)$, 
with parameters $n$, $\a$ and $\b$ integers, 
showing that they are related to 
the unitary irreducible representation of $SU(2,2)$ with quadratic Casimir 
${\cal C}_{SU(2,2)}=-3/2$. 
As  they determine also a base of 
square-integrable functions, the universal enveloping algebra of $su(2,2)$ is homomorphic to the
space of linear operators acting on the $L^2$ functions defined on $(-1,+1) \times {\Z} \times {\Z/2}$.
\end{abstract}

\vskip 1cm


\vfill\eject

\sect{Introduction}\label{intro}

Many attempts have been done to find  a wide but not too inclusive class of functions that can be defined 
``special'', where special means something more that  ``useful'', the old definition of Tur\'an~\cite{andrews1999} (see also Ref.~\cite{berry2001}).

The actual main  line of work for a possible unified theory of special functions is  the Askey scheme  that
is based on the analytical theory of linear differential equations. It
includes as special functions all those functions that are related with hypergeometric functions and their 
$q$-analogues \cite{andrews1999,heckman1994,koekoek2010}. 
In this  approach particular interest is related to the employ of integral transforms, in such a way, that the general hypergeometric function $_p F_q$ can be derived from $_0 F_0 = e^x$ by means of successive applications of Laplace transform and its inverse.
The use of general properties of this  transform and, in particular, of  the convolution theorem enables
also to derive differential identities for the corresponding functions. 

However, a possible alternative point of view of the theory of the basic classes of special functions 
was established by employing considerations that belong to a field of mathematics seemingly quite far from them: the theory of representations of Lie groups. This way was  introduced by Wigner~\cite{wigner1955}  and 
Talman~\cite{talman1968} and
later developed mainly by Miller~\cite{miller1968} and 
Vilenkin and Klimyk~\cite{vilenkin1968,vilenkin1991,vilenkin1995}.
In this line, previous papers by us,  Refs.~\cite{celeghini2013-1,celeghini2013-2}, shown  a direct connection between  some special functions and  well defined Lie groups.

The starting point of our work has been the paradigmatic  example of Hermite functions that are a basis on the Hilbert space of the square integrable functions
defined on the configuration space $\R$. Besides the configuration basis $\{|x\rangle\}_{x\in \R}$,
as well known in the algebraic discussion of the harmonic oscillator,
a discrete basis $\{|n\rangle\}_{n\in \N}$ related to the  Weyl-Heisenberg group $H(1)$ can be considered such that Hermite functions are the transition matrices from one basis to the other~\cite{celeghini2013-1}.
The relevant point is that  this scheme can be generalized
from the Hermite polynomials to all the orthogonal polynomials we have, up to now, considered: Legendre and Laguerre polynomials~\cite{celeghini2013-1}, Associated Legendre polynomials and Spherical Harmonics~\cite{celeghini2013-2} and Jacobi polynomials as we show in this paper.

In the seminal work of 1948 Truesdell introduced indeed the idea that  a sub-class of special functions,  called by him ``familiar special functions",   are defined by means of a  set of  formal properties~\cite{truesdell1948}. 
In his spirit, we proposed in \cite{celeghini2013-1,celeghini2013-2}  a possible definition of a fundamental
sub-class of special functions, we called ``algebraic special functions" (ASF), that  look to be strictly
related to the hypergeometric functions and are constructed starting from algebraic properties. The experience of the cases studied in \cite{celeghini2013-1,celeghini2013-2} leads us to
 consider  the following set of assumptions:
\begin{enumerate}
\item 
A set of recurrence  relations are defined on these ASF that can be associated to a set of ladder operators that span a Lie algebra.
\item 
These ASF support an irreducible representation of this algebra.
\item
A Hilbert space can be constructed on these ASF where these ladder operators have the  hermiticity properties appropriate for constructing a  unitary irreducible representation (UIR) of the associated Lie group.
\item
Second order differential equation that define the ASF can be reconstructed  from all diagonal elements of the universal enveloping algebra (UEA) and, in particular, 
the second orden Casimirs of the subalgebras and  of the whole  algebra.
\end{enumerate} 

From these assumptions, we have that:
\begin{itemize}
\item[\; i)]
Applying the exponential map to ASF different sets of functions can be constructed. If the transformation is unitary another algebraically equivalent basis of the Hilbert space is obtained.
When the transformations are not unitary, as in the case of coherent states, sets with different properties are found (like  overcomplete sets).
\item[ii)]
 The ASF are also a basis of an appropriate set of $L^2$ functions (defined on real spaces) and of an appropriate Hilbert space functions (in complex spaces).  This, combined with the above properties implies that the vector space of the operators operating on $L^2$ (or Hilbert) space functions is homomorphic to the  UEA built on the algebra.
  \end{itemize}

 In \cite{celeghini2013-1}  it has been shown that
 Hermite, Laguerre and Legendre polynomials are ASF such that 
 the Hermite  functions  support a UIR of  the Weyl-Heisenberg group $H(1)$ with Casimir ${\cal C} = 0$, while  Laguerre functions and Legendre polynomials are both bases for the 
 UIR of  $SU(1, 1)$ with ${\cal C}=-1/4$. 
Since
Hermite functions are a  basis of square-integrable  functions  (or wave functions)  defined on the real line, as well as 
Laguerre functions on the semi-line   and  Legendre polynomials on the finite interval~\cite{Cambianis},
all operators of the universal enveloping algebra (UEA) are defined on the appropriate basis functions
and the wave functions  built on them.
In other words, the spaces of linear operators acting on these $L^2$ or Hilbert spaces are homomorphic to the UEA 
(i.e.,  the algebra constructed on the monomials of the Lie algebra generators) of the corresponding Lie algebra.  Of course, this implies that this is true also for the elements of the corresponding Lie group that are contained in the UEA.

 All these properties have been shown to not be restricted to the
above mentioned 1-rank algebras (and groups). Indeed in  \cite{celeghini2013-2}
Associated Legendre polynomials and Spherical Harmonics  are shown to share  the same 
properties. The  underlying  Lie group is in both these
cases $SO(3,2)$ that is of rank 2 like two,  ${l}$ and ${m}$,
are the label parameters of these functions.

Here we discuss a further confirmation of this scheme, now related to
the Jacobi polynomials that also satisfy  the required conditions to be considered ASF and share
the same properties.
Indeed they can be 
associated to well defined ``algebraic Jacobi functions'' that
support a UIR of $SU(2,2)$ i.e.,  a Lie group of rank 3 like three are the parameters, $n, \alpha$ and $\beta$, of the Jacobi polynomials $J_{n}^{(\a,\b)}(x)$. 

The suggestion to consider the Jacobi polynomials as candidates to ASF is related to the well known relation
between Jacobi polynomials with\; $\a = \b$\; and the Associated Legendre Polynomials
\cite{vilenkin1991}:
\be\label{Jaa}
J_{n}^{(\a,\a)}(x)\; =\; (- 2)^\a\; \frac{\Gamma{(n+\a+1)}}{\Gamma{(n+2\a+1)}}\; (1-x^2)^{-\a/2}\;
P^{\a}_{n+\a} (x)
\ee
that shows that
the Jacobi polynomials with $\a=\b$  are ASF related to the Lie group $SO(3,2)$\cite{celeghini2013-2}.
Indeed, since the  relation (\ref{Jaa}) can be read in terms of the notation introduced in   \cite{celeghini2013-2} as
\be\label{ab}
{J}_{n}^{\a,\a}(x)   = (-1)^\a\; {T}_{n+\a}^{\a}(x),
\ee
where ${T}_{l}^{m}(x)$ are the Associated Legendre functions,   the resuls obtained for the Associated Legendre Polynomials ${P}_{l}^{m}(x)$ are easily rewritten for the Jacobi polynomials 
with $\a=\b$.

We present here the generalization of the ladder operators of the Jacobi
polynomials for independent integer values of the labels $\alpha$ and $\beta$.
Obviously as Jacobi polynomials depend from three 
parameters \;$n$\,, $\a$\; and $\b$ we have to look for an algebra of rank three. In the following it
will be shown that this algebra exists and is\; $D_3$ in its real form $su(2,2)$.  

Since the complete construction is complex  it not will be presented here. However, because of the great interest of the Jacobi polynomials in many areas of mathematics and physics, a detailed description  will be published elsewhere.
 
The paper is organized as follows. 
Section~\ref{algebraicJacobifunctions} is devoted to present the main properties of  the algebraic Jacobi functions (AJF) relevant for our discussion.  
In section~\ref{SUA(2)otimesSUB(2)}   we study the symmetries of the  the AJF that keep invariant the parameter $l$ and change $m$ and/or $q$. We prove that these ladder operators determine a $su(2)\oplus su(2)$ algebra,
that  allows us to  build up  
a family of UIR of  the group $SU(2)\otimes SU(2)$ labelled by the parameters $(l,m)$ and $(l,q)$, i.e.  $U^l\otimes U^l$, with $-l\leq m\leq l,\, -l\leq q \leq l$, $l-m\in \N$, $l-q\in \N$ and $2 l\in \N$.
In section~\ref{su11groups} we
construct, by means of four new sets of ladder operators that change the three parameters 
$l, m$ and $q$ in $\pm 1/2$, generating each of them a  $su(1,1)$ algebra to which infinitely many UIR's  of $SU(1,1)$ 
-- supported by the AJF --  are  associated.   
The complete set of ladder operators span a $su(2,2)$ Lie algebra and the AJF generate a UIR of $SU(2,2)$ characterized by the eigenvalue of the quadratic Casimir 
${\cal C}_{SU(2,2)}=-3/2$.
In section~\ref{operatorsl2} the homomorphism between the space of the operators on the $L^2$ space 
and the  UEA of $su(2,2)$ is discussed. 
Finally, in section~\ref{conclusionssection}, some conclusions and comments  are included.

\sect{Algebraic Jacobi functions and their operatorial structure}
\label{algebraicJacobifunctions}

 The Jacobi polynomials are defined in terms of the hypergeometric functions\; $_2F_1$  and 
 Pochhammer's symbol\cite{NIST,luke1969}
\be\label{j}
J_{n}^{(\a,\b)}(x)=\;\frac{(\a+1)_n}{n!}\;\;\;_2F_1\left[-n,1+\a+\b+n;\a+1;\frac{1-x}{2}\right].
\ee

As shown in the following, the set of parameters $\{n, \a, \b\}$ and the same Jacobi polynomials are not directly related to the algebraic structure. 
Thus, consistently with \cite{celeghini2013-1,celeghini2013-2}, we define three new discrete variables and include an $x$-depending factor. Moreover, a peculiar property of the Jacobi polynomials shared with the  Legendre polynomials~ \cite{celeghini2013-1,celeghini2013-2} is that 
the standard form of the matrix elements of the algebra is not connected to an orthonormal basis of
the Hilbert space but to an orthogonal one where
a residual weight is preserved (see eqs. (\ref{ort}) and (\ref{com})).

Hence, we  first introduce in the space of  parameters the  change from the integers 
$(n,\a,\b)$ to   $(l, m, q)$ all togheter integers or half-integers
\[
 l:=n+\frac{\a+\b}{2}\,, \qquad  m:=\frac{\a+\b}{2}\,,\qquad  q:=\frac{\a-\b}{2}\,,
\]
or, equivalently,
\[
 n=l - m, \qquad  \a= m+q,\qquad  \b= m-q\; .
\]
Then,  in a second and final step,  we include a $x$-depending factor
\[
\sqrt{\frac{\G (l+ m+1)\,\G (l-m+1)}{2^{2m}\,
\G (l+q+1)\,\G (l-q+1)}}\,\,
(1-x)^{\frac{m+q}{2}}\, (1+x)^{\frac{m-q}{2}} .
\]
Thus,  the fundamental objects of this paper,  that we call 
``algebraic Jacobi functions'' (AJF), have the final  form
\[\label{AJF}
{\cal J}_l^{m,q}(x):=
\sqrt{\frac{\G (l+ m+1)\,\G (l-m+1)}{2^{2m}\,
\G (l+q+1)\,\G (l-q+1)}}\,\,
(1-x)^{\frac{m+q}{2}}\, (1+x)^{\frac{m-q}{2}} \,J_{l-m}^{(m+q,m-q)}(x)\;. 
\]

 These new objects reveal  additional symmetries hidden inside the Jacobi polynomials  (\ref{j}) to be added to the well known
$J_{n}^{(\a,\b)}(x)= (-1)^n J_{n}^{(\b,\a)}(-x)$\,. For instance,
\be\label{symmj}
\begin{array}{lll}
{\cal J}_l^{m,q}(x)={\cal J}_l^{q,m}(x),\\[0.3cm]
{\cal J}_l^{m,q}(x)=(-1)^{m+q} {\cal J}_l^{-m,-q}(x),  \\[0.3cm]
{\cal J}_l^{m,q}(x)=(-1)^{l-q} {\cal J}_l^{-m,q}(-x),  \\[0.3cm]
{\cal J}_l^{m,q}(x)=(-1)^{l-m} {\cal J}_l^{m,-q}(-x)\,.
\end{array}
\ee
Furthermore these functions verify the normalization relations for $m$ and $q$ fixed
\be\begin{array}{lll} \label{com}
\ds\int_{-1}^{1}\,{\cal J}_l^{m,q}(x)\,(l+1/2)\, {\cal J}_{l'}^{m,q}(x) &=&
\delta_{l\, l'},\\[0.4cm]
\ds \sum_{l={\it sup}(|m|, |q|)}^\infty  {\cal J}_l^{m,q}(x) \left(l+1/2\right) {\cal J}_l^{m,q}(y) &=& \delta(x-y). 
\end{array}\ee
similar to the ones imposed in \cite{celeghini2013-1} to the Legendre polynomials and in \cite{celeghini2013-2} to the
associated Legendre polynomials.
Note again that they are orthogonal but, like in the SO(3,2) case, orthonormal only up to the factor $l+1/2$. 

The Jacobi equation 
\[
\left[ (1-x^2) \frac{d^2}{dx^2} -((\a+\b+2)x+(\a-\b)) \frac{d}{dx} + n(n+\a+\b+1)\right]\,
J_n^{(\a,\b)}(x)=0 
\]
rewritten in terms of these new functions ${\cal J}_l^{m,q}(x)$ and  of the new parameters $l,m$ and $q$ becomes 
\be\label{jacobiequation}
\left[-(1-x^2)\,\frac{d^2}{dx^2}+2\,x\frac{d}{dx}+ 
\frac{2\,m\,q\,x+m^2+q^2}{1-x^2}- l(l+1)\right] \,{\cal J}_l^{m,q}(x)=0\; .
\ee
It is worthy noticing  the symmetry under the interchange $m \Leftrightarrow q$ in the expressions  \eqref{symmj} and \eqref{jacobiequation}.

In the spirit of  \cite{celeghini2013-1,celeghini2013-2}, the starting point for the construction of the algebra associated to these algebraic Jacobi functions is now
the construction of the rising/lowering differential operators that allow to obtain  from each AJF
the contiguous ones differing by 1 or 1/2 in the value of the discrete variables $l,m$ and  $q$. The fundamental limitation of this approach is  
that the problem has been considered  from the point of view of differential equations where the indices 
are considered as parameters \cite{miller1968}. The dependence of the formulas from the 
indices in iterated applications must thus be introduced by hand.
This problem has been taken into account in  \cite{celeghini2013-1} 
where a consistent vector space framework (where the indices are related to discrete operators) was introduced
to allow the iterated use of recurrence formulas by means of operators. 
The parameters involved are thus eigenvalues of certain discrete operators acting on the space of the AJF.

Thus, in order to display the complete operator structure on the set 
$\{{\cal J}_l^{m,q}(x)\}$  we introduce,
in consistency with the quantum theory approach, not only the 
operators $X$ and $D_x$ of the configuration space, such that 
\[
X\, f(x) = x\, f(x) ,\qquad D_x\, f(x) = f'(x),\qquad [X,D_x]= -1\, ,
\]
but  also three other operators $L$, $M$ and $Q$ such that 
\be\label{lmq}
L\; {\cal J}_l^{m,q}(x) = l\; {\cal J}_l^{m,q}(x) \quad\; M\, {\cal J}_l^{m,q}(x) = \, m\; {\cal J}_l^{m,q}(x)\,
\quad\; Q\, {\cal J}_l^{m,q}(x) = \, q\; {\cal J}_l^{m,q}(x),  
\ee
i.e. diagonal on the Jacobi functions and, thus, commuting between them 
\[
[ L, M ] = [ L, Q ] = [ M, Q ] = 0 .
\]
Hence, when we shall consider the whole algebra, all of them will belong to the corresponding
Cartan subalgebra. It will be proved later that the  obtained representation of the corresponding group ($SU(2,2)$) is  unitary if the variables $l, m, q$ are such that 
\[
l\geq |m|,\;\; \; l\geq|q|, \qquad   2 l,\;\; l-m,\;\; l-q \;\in \N\;.
\]


\sect{$SU_A(2)\otimes SU_B(2)$  for Jacobi functions with $\Delta l = 0$}\label{SUA(2)otimesSUB(2)}

We have now to introduce the action of the ladder operators  on the set of the Jacobi functions, $\{{\cal J}_l^{m,q}(x)\}$, as differential-difference relations. 

We start from the differential-difference 
equations and the difference equations verified by the Jacobi functions, a  complete list of which can be found in
Refs.~\cite{NIST,luke1969,abramowitz1972}.
The  procedure is  laborious, and, as we said before, it will be
reproduced  in a more detailed way  in a following paper. Here we only sketch the simplest examples in order to enlighten the procedure.
 Let us consider the equations (18.9.15) and  (18.9.16)  of Ref.~\cite{NIST}
\[\begin{array}{rll}
\ds \frac{d}{dx} P_{n}^{(\alpha , \beta)}(x) &=& \frac{1}{2}(n+\alpha+\beta+1)\,P_{n-1}^{(\alpha+1,\beta+1)}(x)\;,
\\[0.3cm]
\ds\frac{d}{dx}\left[(1-x)^\alpha (1+x)^\beta P_{n}^{(\alpha , \beta)}(x)\right] &=& -2(n+1) (1-x)^{\alpha-1}
(1+x)^{\beta-1}\,P_{n+1}^{(\alpha-1, \beta-1)}(x)
\end{array}\] 
which are far to be symmetric.
But if we rewrite them  in terms of the algebraic Jacobi functions  
${\cal J}_l^{m,q}(x)$   they allow to define the operators 
\be\label{Adiferential} 
A_{\pm} :=\; \pm \,\sqrt{1-X^2}\, D_x\,
+\, \frac{1}{\sqrt{1-X^2}}\; (X M+Q)
\ee
that act in the following way 
\be\label{Anodiferential}
A_\pm\; {\cal J}_l^{m,q}(x) = \sqrt{(l\mp m)\,(l \pm m+1})\;
\,{\cal J}_{l}^{m\pm 1,\, q}(x) .
\ee

The operators (\ref{Adiferential})  are a generalization for $q\neq 0$ of the
operators $J_\pm$ introduced  in Ref.~\cite{celeghini2013-2}  
for the Associated Legendre functions related by eq.~\eqref{ab} 
to the
Jacobi functions  with $q=0$. Moreover 
eqs.~(\ref{Anodiferential}), that are independent from $q$,
coincide with eqs. (2.11) and  (2.12) of Ref.~\cite{celeghini2013-2}.  

Taking into account the action of the operators $A_\pm$ and $M$ on the Jacobi functions, eqs. \eqref{Anodiferential} and \eqref{lmq}, respectively, and defining  $A_3:= M$ it is easy to check that 
$A_\pm$ and $A_3$ close a $su(2)$ algebra,  denoted in the following $su_A(2)$,
\[\label{suA2}
[A_3,A_\pm]=\pm A_\pm \qquad [A_+,A_-]=2 A_3, 
\]
and  commute with  $L$ and $Q$
\be\label{Alq} 
[L, A_\pm]=0 ,\qquad [Q,A_\pm]=0 ,\qquad [L,A_3]=0,\qquad [Q,A_3]=0 .
\ee 

Note that from \eqref{Anodiferential} and \eqref{lmq} the  Jacobi functions $\{{\cal J}_l^{m,q}(x)\}$ such that  $2 l \in \N$,
$l-m \in \N$ and  $-l\leq m\leq l  $  support  the  $(2 l+1)$-dimensional UIR of the
Lie group  $SU_A(2)$ independently from the value of $q$.

Like in \cite{celeghini2013-2}, starting from 
the differential realization  \eqref{Adiferential}
of the $A_\pm$ operators 
 we recover the Jacobi differential equation \eqref{jacobiequation}  from the Casimir of $su_A(2)$ $({\cal C}_A)$, i.e.,
\be\label{eqcasimirA}
 \left[{\cal C}_A-L(L+1)\right]\;{\cal J}_l^{m,q}(x)\equiv 
 \left[A_3^2+\frac12\{A_+,A_-\}-L(L+1)\right]\,{\cal J}_l^{m,q}(x) = 0\;.
 \ee
 Effectively, eq.~\eqref{eqcasimirA} reproduces the operatorial form of \eqref{jacobiequation}, i.e.
\be\label{jacobiequationoperator}
[-(1-X^2) D^2_x + 2 X D_x + \frac{1}{1-X^2} (2 X M Q +M^2 + Q^2) - L(L+1)] {\cal J}_l^{m,q}(x) = 0 ,
\ee
On the other hand  we can make use of
the factorization method \cite{schrodinger,infeld-hull1951},  that relates  second order differential equations to  recurrence formulae written in terms of first order derivatives in such a way that  the application of the first operator modifies the values of the parameters of the second one. 
Taking into account this fact, by means of iterate application of   \eqref{Anodiferential}
we obtain that  the two equations
 \be \begin{array}{l}\label{jacobinodiferential1}
\left[A_+\,A_- -(L+M)\,(L-M+1)\right]\; {\cal J}_l^{m,q}(x)=0\;,
\\[0.3cm]
\left[A_-\,A_+ -(L-M)\,(L+M+1)\right]\; {\cal J}_l^{m,q}(x)=0\;,
\end{array}\ee
reproduces the Jacobi equation \eqref{jacobiequation} in operator form \eqref{jacobiequationoperator}.
Equations \eqref{eqcasimirA}, \eqref{jacobiequationoperator} and    \eqref{jacobinodiferential1}  are particular cases of a general rule:  the defining Jacobi equation can be recovered from the Casimir operator   of any  algebra and sub-algebra involved acting in 
${J_l^{m,q}}$ as well as from any diagonal product of ladder operators.

Now the symmetry under the interchange\, $m \Leftrightarrow q$\, in ${\cal J}_l^{m,q}(x)$ exhibited 
in eqs.~(\ref{symmj}) and \eqref{jacobiequation} allows to define two  new operators $B_\pm$  from  
$A_\pm$ 
 by means of the exchange 
\be\label{absymm}
B_\pm (X,D_x,M,Q) = A_\pm (X,D_x,Q,M)\ee
Thus,
\be \label{Bdiferential}
B_{\pm} := \pm \,\sqrt{1-X^2}\, D_x+\frac{1}{\sqrt{1-X^2}} \,   (X Q+M)\,,
\ee
such that their action on the Jacobi functions  is
\be \label{Bnodiferential}
B_\pm\,{\cal J}_l^{m,n}(x) = \sqrt{(l\mp q)\,(l \pm q+1})\; 
\,{\cal J}_{l}^{m,q\pm 1}(x) .
\ee

Obviously also the operators $B_\pm$ and $B_3:=Q$ close a $su(2)$ algebra   we denote $su_B(2)$
\[
[B_3,B_\pm]=\pm B_\pm \qquad [B_+,B_-]=2 B_3 ,
\]
and the Jacobi functions $\{{\cal J}_l^{m,q}(x)\}$  with $2l\in \N$, $l-q\in \N$ and $-l\leq q \leq l$ close the  $(2 l+1)$-dimensional UIR of the Lie group  $SU(2)_B$ independently from the value of $m$.

Moreover  similarly to the commutation relations  between operators $A_\pm, A_3$ and $L,Q$ \eqref{Alq} we have now
\[ 
[L, B_\pm]=0 \qquad [M,B_\pm]=0, \qquad [L,B_3]=0 , \qquad [M,B_3]=0 .
\] 

Again we can recover the Jacobi equation  \eqref{jacobiequation} from the Casimir of the algebra $su_B(2)$ 
\[
\left[{\cal C}_B-L(L+1)\right] {\cal J}_l^{m,q}(x) = 
\left[  B_3^2 +      \frac12\{B_+,B_-\}-L(L+1)\right] {\cal J}_l^{m,q}(x)=0
\]
as well as from the expressions 
\[ \begin{array}{l}
\left[B_+\,B_- -(L+Q)\,(L-Q+1)\right]\; {\cal J}_l^{m,q}(x)=0\;,
\\[0.3cm]\left[B_-\,B_+ -(L-Q)\,(L+Q+1)\right]\; {\cal J}_l^{m,q}(x)=0\;.
\end{array}\]

\begin{figure}
\centerline{\psfig{figure=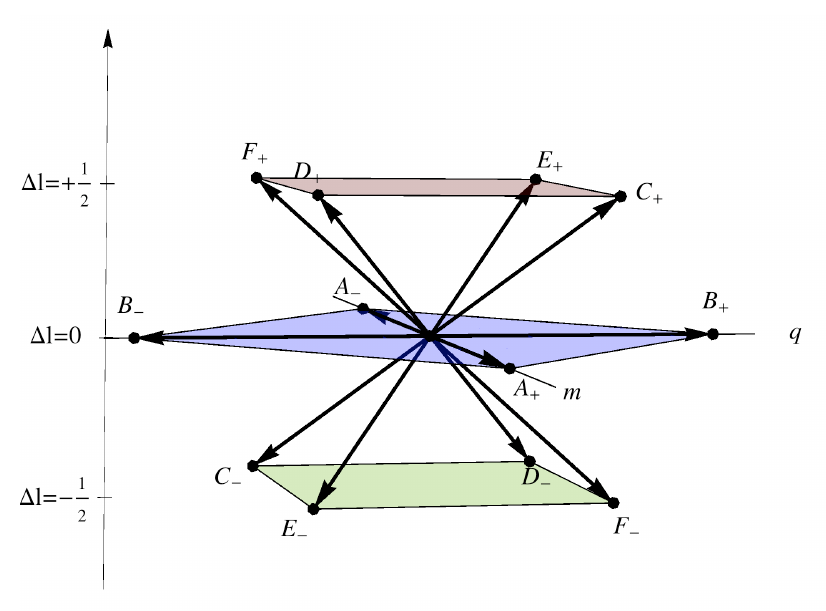,height=8.0cm}}
\caption{\small Action of the ladder operators  on the Jacobi functions ${\cal J}_l^{(m,q)}(x)$ represented by the triplets $(l,m,q)$. The planes displayed correspond to the pairs $(m,q)$, the parameter $l$ (or $\Delta l$) is represented in the vertical axis.} \label{fig_1}
\end{figure}

A more complex algebraic scheme  appears 
in common applications of the operators $A_\pm$ and $B_\pm$.  As the operators $A_\pm, A_3$  commute with $B_\pm,B_3$,  the algebraic structure is the  direct sum of Lie algebras
\[
su_A(2)\oplus su_B(2) .
\]
A new symmetry of the  AJF emerges that moves $m$ and $q$ without
changing $l$. For $l, m, q$ integer and half-integer formulae 
\eqref{Anodiferential}, \eqref{Bnodiferential} and \eqref{lmq}  are the well known expressions for the infinitesimal generators
 of the group $SU_A(2)\otimes SU_B(2)$.
 The Jacobi functions 
${\cal J}_l^{m,q}(x)$ for fixed $l$ and $-l \leq m\leq l$, $-l \leq q\leq l$ determine a UIR  of this group. As we mention previously, the structure of the Hilbert space of the AJF and the hermiticity of
the generators will be discussed in Sect~\ref{operatorsl2}. In Fig.~\ref{fig_1}   the action of the operators $A_\pm,B_\pm$ on the parameters $(l,m,q)$ that label the Jacobi functions  corresponds to the plane $\Delta l=0$. 

\sect{Other ladder operators inside algebraic Jacobi functions and $su(1,1)$ representations}\label{su11groups}

We mention before that  many difference and differential-difference relations for Jacobi polynomials are 
known~\cite{luke1969, NIST}.  
Starting from them a su(2,2) Lie algebra can be constructed.  It 
has fifteen infinitesimal generators, three of them are Cartan generators (for instance, $L, M$ and $Q$ 
or three independent
linear combinations of them). We have  four generators ($A_\pm$ \eqref{Adiferential} and $B_\pm$ \eqref{Bdiferential} ) that commute with 
$L$, hence  we need eight non-diagonal
operators more. They are in differential form:
\be\label{Differential8}
\displaystyle
\begin{array}{l}
C_+:=\,\dfrac{(1+X)\sqrt{1-X}}{\sqrt{2}}\, D_x - \dfrac{1}{\sqrt{2(1-X)}} \left(X\,(L+1)-(L+1+M+Q)\right),
\\[0.5cm]
C_-:=-\,
\dfrac{(1+X)\sqrt{1-X}}{\sqrt{2}}\, D_x - \dfrac{1}{\sqrt{2(1-X)}}\; \left(X\, L- (L+M+Q)\right),
\\[0.5cm]
D_+:=-\,
\dfrac{(1-X)\sqrt{1+X}}{\sqrt{2}}\, D_x + \dfrac{1}{\sqrt{2\,(1+X)}}\; \left(X (L+1)+ (L+1+M-Q)\right),
\\[0.5cm]
D_-:=+\,
\dfrac{(1-X)\sqrt{1+X}}{\sqrt{2}} \, D_x + \dfrac{1}{\sqrt{2\,(1+X)}}\; \left(X\,L + (L+M-Q)\right),
\\[0.5cm]
E_+:=-\,
\dfrac{(1-X)\sqrt{1+X}}{\sqrt{2}}\, D_x+
\dfrac{1}{\sqrt{2\,(1+X)}}\left( X (L+1)+(L+1-M+Q) \right),
\\[0.5cm]
E_-:=+\,\dfrac{(1-X)\sqrt{1+X}}{\sqrt{2}}\, D_x+\dfrac{1}{\sqrt{2(1+X)}}\; \left( X L+(L-M+Q) \right),
\\[0.5cm]
F_+:=-\,
\dfrac{(1+X)\sqrt{1-X}}{\sqrt{2}}\, D_x+
\dfrac{1}{\sqrt{2\,(1-X)}}\left( X (L+1)-(L+1-M-Q) \right),
\\[0.5cm]
F_-:=+\,
\dfrac{(1+X)\sqrt{1-X}}{\sqrt{2}}\, D_x+\dfrac{1}{\sqrt{2\,(1-X)}}\,
 \left( X L-(L-M-Q) \right).
\end{array}
\ee
All these differential operators act on the space $\{{\cal J}_l^{m,q}\}$ for $l,m,q$ integer and
half-integer such that $l\ge |m|,|q|$. The explicit form of its action is: 
\be\label{Nodifferential8}
\displaystyle
\begin{array}{rcl}
C_+\,{\cal J}_l^{m,q}(x)&=&\sqrt{(l+m+1)(l+q+1)}\; {\cal J}_{l+1/2}^{m+1/2,\;q+1/2}(x),\\[0.4cm]
C_-\,{\cal J}_n^{m,q}(x)&=&\sqrt{(l+m)(l+q)}\; {\cal J}_{l-1/2}^{m-1/2,\;q-1/2}(x),\\[0.4cm]
D_+\,{\cal J}_l^{m,q}(x)&=&\sqrt{(l+m+1)(l-q+1)}\; {\cal J}_{l+1/2}^{m+1/2,\;q-1/2}(x)\\\\
D_-\,{\cal J}_l^{m,q}(x)&=&\sqrt{(l+m)(l-q)}\; {\cal J}_{l-1/2}^{m-1/2,\,q+1/2}(x,)\\[0.4cm]
E_+\,{\cal J}_l^{m,q}(x)&=&\sqrt{(l-m+1)\,(l+q+1)}
\;{\cal J}_{l+1/2}^{m-1/2,\;q+1/2}(x),\\[0.4cm]
E_-\,{\cal J}_l^{m,q}(x)&=&\sqrt{(l-m)\,(l+q)}\;
{\cal J}_{l-1/2}^{m+1/2,\;q-1/2}(x),\\[0.4cm]
F_+\,{\cal J}_l^{m,q}(x)&=&\sqrt{(l-m+1)\,(l-q+1)}\;
\,{\cal J}_{l+1/2}^{m-1/2,\,q-1/2}(x),\\[0.4cm]
F_-\,{\cal J}_l^{m,q}(x)&=&\sqrt{(l-m)\,(l-q)}\;
{\cal J}_{l-1/2}^{m+1/2,\;q+1/2}(x).
\end{array}
\ee
From \eqref{Nodifferential8} (analogously to what happen with 
$A_\pm$ and $B_\pm$) it is obvious that
\[
C_\pm^\dagger = C_\mp,\qquad 
 D_\pm^\dagger = D_\mp,\qquad
  E_\pm^\dagger = E_\mp,\qquad
   F_\pm^\dagger = F_\mp,
\]
i.e. all these rising/lowering  operators have the hermiticity properties required by the representation to be unitary.
Note that   the operators defined in  \eqref{Differential8}  change the parameters $(l,m,q)\to (l\pm 1/2,\,m\pm 1/2,\,q\pm 1/2)$.  More precisely,
 $X_+:l\to l+1/2$ and  $X_-:l\to l-1/2$. In Fig.~\ref{fig_1}   the action of these operators   on the labels $(l,m,q)$ corresponds to the planes $\Delta l=\pm1/2$. 

From the eqs.~\eqref{Differential8}   the following relations among these new operators are easily stated
\be\begin{array}{llr}\label{weylsym}
 D_\pm (X,D_x,M,Q)  &=  &C_\pm (-X,-D_x,M,-Q),\\[0.3cm]
E_\pm (X,D_x,M,Q)  &=  &C_\pm (-X,-D_x,-M,Q),\\[0.3cm]
F_\pm (X,D_x,M,Q)  &=  &- C_\pm (X,D_x,-M,-Q)
\end{array}\ee
Because of the symmetries \eqref{weylsym} we can  only discuss the operators 
$C_\pm$. Along this section we will see that this symmetry of interchange between the ladder operators (\eqref{absymm} and   \eqref{weylsym})
 can be identify   with  the Weyl symmetry acting on  the roots of the simple Lie algebra that they span ($su(2,2)$ in this case).

Taking thus into account, like in the previous cases, the action of the operators $C_\pm$ and $L, M, Q$ on the Jacobi functions, 
eqs. \eqref{Differential8} and \eqref{lmq}, respectively, we get that 
\be\label{suC11}
 [C_+,C_-]= -2 C_3,  \qquad   [C_3,C_\pm]=\pm C_\pm 
\ee
where 
\be \label{c3spectrum}
C_3:= L+ \frac{1}{2}(M+Q)+\frac12.
\ee 
Hence $\langle C_\pm,C_3\rangle$ close a $su(1,1)$ algebra that we will denote, as usual,  $su_C(1,1)$.

As in the  cases of the operators $A_\pm$ and $B_\pm$, we recover the Jacobi differential equation \eqref{jacobiequation} up to a nonvanishing factor
using the differential realization  \eqref{Differential8}
of the operators $C_\pm$  
  from the Casimir of $su_C(1,1)$ 
  \be\label{casimirsuC11}
  {\cal C}_C{\cal J}_l^{m,q}(x)\equiv\left[C_3^2-\frac12\{C_+,C_-\}\right]{\cal J}_l^{m,q}(x)=\frac{1}{4}\left[(m+q)^2-1\right]{\cal J}_l^{m,q}(x).
  \ee
 So 
\be\begin{array}{l}\label{eqcasimirC}
 \ds\left[{\cal C}_C-\frac{(M+Q)^2-1}{4}\right]\;{\cal J}_l^{m,q}(x)\\[0.5cm]
\qquad \equiv 
 \left[C_3^2-\frac12\{C_+,C_-\}-\frac{1}{4}\left((M+Q)^2-1\right)\right]\,{\cal J}_l^{m,q}(x) =0 .
\end{array}\ee
Analogously  the factorization method from a diagonal  product of  two ladder operators  gives
\be\begin{array}{l}\label{Cjacobinodiferential1}
\left[C_+\,C_- -(L+M)\,(L+Q)\right]\; {\cal J}_l^{m,q}(x) =0 ,
 \\[0.4cm]
\left[C_-\,C_+ -(L+1+M)\,(L+1+Q)\right]\; {\cal J}_l^{m,q}(x)=0 , 
 \end{array}\ee
and all the three eqs.~\eqref{eqcasimirC} and \eqref{Cjacobinodiferential1} allow us to  recover the Jacobi equation \eqref{jacobiequationoperator}.

From \eqref{eqcasimirC} we see that since $(m+q) = 0,\pm 1,\pm 2,\pm 3,\cdots $ the IR of su(1, 1) with 
${\cal C}_C =( (m+q)^2-1)/4 = -1/4, 0, 3/4,2, 15/4, \cdots$ are obtained. Moreover the  unitarity of these IR comes from the  the fact that $C^\dagger_\pm=C_\mp$
\cite{Bargmann47}.  The spectrum
of the operator $C_3$  is given by \eqref{c3spectrum} and, since  $l\geq|m|$ and $l\geq |q|$, has the eigenvalues 
$\text{sup}(|m|,|q|),\text{sup}(|m|,|q|)+1/2,\text{sup}(|m|,|q|)+1, \text{sup}(|m|,|q|)|+3/2, \cdots$
Hence,  the set of AJF 
  supports many
 infinite dimensional UIR, of $SU(1, 1)$ of the discrete series for $SU_C(1,1)$.

Similar results can be found for the other  ladder operators $D\pm,
E\pm,F\pm$ provide that we do the corresponding exchange displayed in \eqref{weylsym} on all the eqs.~\eqref{suC11}, \eqref{casimirsuC11} \eqref{Cjacobinodiferential1}, \eqref{eqcasimirC} and \eqref{Cjacobinodiferential1}.

\sect{The complete symmetry group of  $\{{\cal J}_l^{m,q}(x)\}$: $SU(2,2)$}\label{so3section}

If one represents the action of the twelve operators $A_\pm ,A_\pm , C_\pm , D_\pm , E_\pm ,
F_\pm$, that we have defined in previous sections, we obtain Fig.~\ref{fig_1}.
To obtain the root system of the simple Lie algebra $A_3 \equiv D_3$ we have only simply to add three points in the origin
corresponding to the elements  $L, M$ and  $Q$ of the Cartan subalgebra.

\subsect{Commutation relations}

The Lie commutators of the generators $A_\pm,B_\pm,C_\pm,D_\pm,E_\pm,F_\pm,M,  Q,L$ are 
\[\begin{array}{llll}
[L,A_\pm]=0,\quad &
[L,M]=0,\quad &[L,B_\pm]=0,\quad &[L,Q]=0,
\\[0.4cm]
[L,C_\pm]=\pm \frac12 \,C_\pm,\quad &
[L,D_\pm]=\pm \frac12 \,D_\pm,\quad &[L,E_\pm]=\pm \frac12 \,E_\pm,\quad &[L,F_\pm]=\pm \frac12 \,F_\pm,
\\[0.4cm]
  [M,B_\pm]=0,\quad & [M,Q]=0,&&
\\[0.4cm]
[M,C_\pm]=\pm\frac12\, C_\pm,\quad &[M,D_\pm]=\pm\frac12\, D_\pm,\quad &[M,E_\pm]=\mp\frac12\, E_\pm,\quad &[M,F_\pm]=\mp\frac12\, F_\pm,
\\[0.4cm]
  [Q,A_\pm]=0,\quad &&&
\\[0.4cm]
[Q,C_\pm]=\pm \frac12 \,C_\pm,\quad &[Q,D_\pm]=\mp \frac12 \,D_\pm,\quad &[Q,E_\pm]=\pm \frac12 \,E_\pm,\quad &[Q,F_\pm]=\mp \frac12 \,F_\pm,
\\[0.4cm]
[A_+,A_-]=2 A_3,\quad &[A_3,A_\pm]=\pm A_\pm,  &(A_3=M) , & 
\\[0.4cm]
[B_+,B_-]=2 B_3,\quad &[B_3,B_\pm]=\pm B_\pm, &(B_3=Q), &
\\[0.4cm]
[C_+,C_-]=-2 C_3,\quad & [C_3,C_\pm]=\pm C_\pm, &(C_3= L+ \frac{1}{2}(M+Q)+\frac12) ,&
\\[0.4cm]
[D_+,D_-]=-2 D_3,\quad &[D_3,D_\pm]=\pm D_\pm, &(D_3= L+ \frac{1}{2}(M-Q)+\frac12),&
\\[0.4cm]
[E_+,E_-]=-2 E_3,\quad & [E_3,E_\pm]=\pm E_\pm, &(E_3= L+ \frac{1}{2}(-M+Q)+\frac12),&
\\[0.4cm]
[F_+,F_-]=-2 F_3,\quad & [F_3,F_\pm]=\pm F_\pm, &(F_3= L-\frac{1}{2}(M+Q)+\frac12),&
\\[0.4cm]
[A_\pm,B_\pm]=0,\quad &[A_\pm,B_\mp]=0,\quad & &
\\[0.4cm]
[A_\pm,C_\pm]=0,\quad &[A_\pm,C_\mp]=\pm E_\mp,\quad &[A_\pm,D_\pm]=0,\quad & [A_\pm,D_\mp]=\mp F_\mp,
\\[0.4cm]
[A_\pm,E_\pm]=\pm C_\pm,\quad &[A_\pm,E_\mp]=0,\quad &[A_\pm,F_\pm]=D_\pm,
\quad &[A_\pm,F_\mp]=0,
\\[0.4cm]
[B_\pm,C_\pm]=0,\quad &[B_\pm,C_\mp]=\mp D_\mp,\quad &[ B_\pm,D_\pm]=\pm C_\pm,
\quad &[B_\pm,D_\mp]=0,
\\[0.4cm]
[B_\pm,E_\pm]=0,\quad &[B_\pm,E_\mp]=\mp F_\mp ,\quad &[B_\pm,F_\pm]=\pm E_\pm,
\quad &[B_\pm,F_\mp]=0,
\\[0.4cm]
[C_\pm,D_\pm]=0,
\quad &[C_\pm,D_\mp]=\mp B_\pm,\quad
&[C_\pm,E_\pm]=0,
\quad &[C_\pm,E_\mp]=\mp A_\pm,
\\[0.4cm]
[C_\pm,F_\pm]=0,
\quad &[C_\pm,F_\mp]=0, &&
\\[0.4cm]
[D_\pm,E_\pm]=0,
\quad &[D_\pm,E_\mp]=0,\quad
&[D_\pm,F_\pm]=0,
\quad &[D_\pm,F_\mp]=\mp A_\pm,
\\[0.4cm]
 [E_\pm,F_\pm]=0,\quad & [E_\pm,F_\mp]=\mp B_\pm . &&
\end{array}\]

\subsect{Casimir of $su(2,2)$}

The quadratic  Casimir of $su(2,2)$ has the form 
\[\begin{array}{ll}
{\cal C}_{su(2,2)}&=\ds
\frac12\left(\{A_+,A_-\}+\{B_+,~B_-\}-\{C_+,C_-\}-\{D_+,D_-\}  -\{E_+,E_-\}-\{F_+,F_-\}\right)\\[0.3cm]
&\qquad\qquad\ds+\,
\frac{1}{2}\,\left( A_3^2 + B_3^2 + C_3^2 + D_3^2 + E_3^2+ F_3^2 \right)\\[0.4cm]
&=\ds
\frac12\left(\{A_+,A_-\}+\{B_+,~B_-\}-\{C_+,C_-\}-\{D_+,D_-\}  -\{E_+,E_-\}-\{F_+,F_-\}\right)\\[0.3cm]
&\qquad\qquad\ds+\,2 L(L+1)+M^2+Q^2+\frac12\\[0.4cm]
&\ds \equiv -\frac{3}{2}
\end{array}\]
From it and taking into account the differential realization  of the operators involved,
\eqref{Adiferential},  \eqref{Bdiferential} and \eqref{Differential8},  we recover again the Jacobi equation  
\eqref{jacobiequation}.

\begin{figure}
\centerline{\psfig{figure=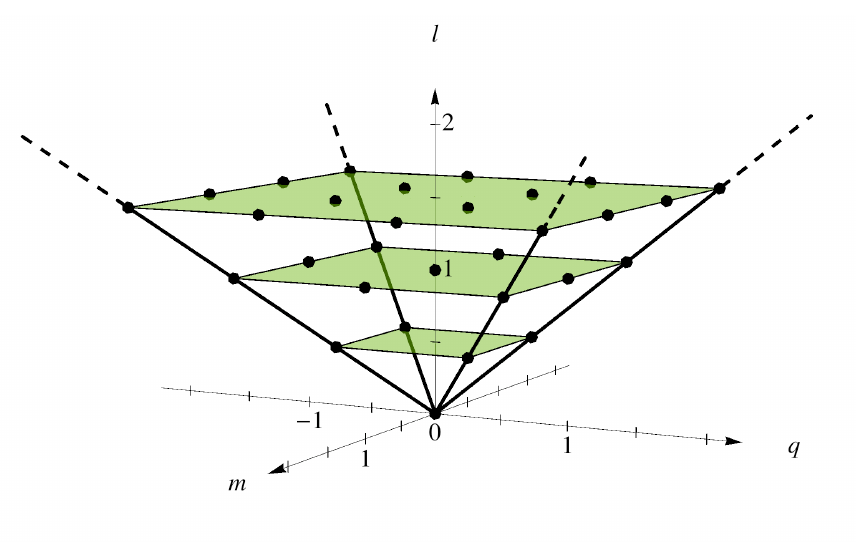,height=8.0cm}}
\caption{\small IR of $su(2,2)$ supported by the AJF ${\cal J}_l^{m,q}(x)$ represented by the black points. The horizontal planes correspond to IR of $su_A(2)\oplus su_B(2)$.} \label{fig_3}
\end{figure}

Hence, the AJF support a UIR of the group $SU(2,2)$ with the value -3/2 of the Casimir 
${\cal C}_{su(2,2)}$ (see Fig.~\ref{fig_3}). Also, as we have seen along the previous sections, the Jacobi equations is recovered form the Casimir of any of the 3-dimensional subalgebras of  ${su(2,2)}$. 

A peculiar property of the Algebraic Special Functions, that perhaps can be assumed as their definition, seems to be
 that, using the fundamental second order differential equation, all diagonal elements of the UEA can be found to be equivalent to it while all non-diagonal elements can be written as first order differential operators.


\sect{Operators on $L^2$ spaces and UEA}
\label{operatorsl2}

We have shown in the preceding sections that the algebraic Jacobi functions  ${\cal J}_l^{mq}$ with $l, m, q$ all
together integer or half-integer  
are a basis of a IR of $su(2,2)$.
They are a basis of the $L^2$ functions defined on $\E\times \Z \times \Z/2$ , where $\E=(-1,1)\subset\R$ and
\[
\Z/2 := \{\cdots,-3/2,-1, -1/2, 0, 1/2, 1, 3/2, \cdots\}
\]
is related to $m$ and $ \Z$ to $m-q$.
Hence, the property that the vector 
space of linear operators acting on them are homomorphic to the UEA of $su(2,2)$ is extended to these $L^2$ functions.  The results presented in \cite{celeghini2013-1, celeghini2013-2} 
for the one-variable and two-variable square-integrable functions, respectively, can be extended to the Jacobi case.

Thus, let us consider the space of square functions defined on ${\E}\times\Z \times \Z/2$,  
${L}^2(\E,\Z,\Z/2)$, which is the  direct sum of 
the Hilbert spaces with $m$ and $q$  fixed, ${L}^2(\E,m,q)$:
\[
{L}^2(\E,\Z,\Z/2) = \bigcup_{m-q\in\Z} \; \bigcup_{q\in\Z/2}  \;{L}^2(\E,m,q).
\]

A basis for ${L}^2(\E,\Z,\Z/2)$ is $\{|x, m,q\rangle\}$ ($-1 < x <1, m,q \in \Z/2, m-q\in\Z$). Orthonormality and 
completeness relations are, respectively,
\[
\langle x,m,q | x', m',q' \rangle\; =\; \delta(x-x')\, \delta_{m\, m'}\, \delta_{q\,q'} ,\qquad
\sum_{m,q }\, \int_{-1}^{+1} dx\, |x,m,q\rangle\,  \langle x,m,q|\; =\; {\cal I}.
\]

As the set  $\{{\cal J}_l^{m,q}(x)\}$ satisfy eqs.~(\ref{ort}) and (\ref{com})  
we can now define inside the Hilbert space a new basis  $\{|l,m,q\rangle\}$ with $l, m,q\in\Z/2,\; l \geq |m|, \; l \geq |q|,\;l-m\in\N,\;l-q\in\N$
\[
|l,m,q\rangle := \int_{-1}^{+1} |x,m,q\rangle \,\sqrt{l+1/2}\; \,{\cal J}_l^{m,q}(x) \,dx\;.
\]
such that
\[
\langle l,m, q' | l', m', q' \rangle\; =\; \delta_{l\,l'}\, \delta_{m\, m'}\, \delta_{q,\, q'},
\qquad
\sum_{ l, m, q}  |l,m, q\rangle\;   \langle l,m, q|\; =\; {\cal I} ,
\]
where  the ${\cal J}_l^{m,q}(x)$ play  the role of transition matrices: 
\[
{\cal J}_l^{m,q}(x) =\frac{1}{\sqrt{l+1/2}}\;  \langle x,m,q|l,m,q\rangle = 
\frac{1}{\sqrt{l+1/2}}\;\langle l,m,q|x,m,q\rangle \;. 
\]
This  transition matrix role of the algebraic Jacobi functions $\{{\cal J}_l^{m,q}(x)\}$ reflects the fact that the generators can be seen as differential operators on the variable space 
\eqref{Adiferential}, \eqref{Bdiferential} and \eqref{Differential8} or algebraic operators in the spaces of labels \eqref{Anodiferential}, \eqref{Bnodiferential} and \eqref{Nodifferential8} allowing to make explicit the Lie algebra structure in contraposition to previous works~\cite{miller1968,vilenkin1968,vilenkin1991}.

In  analogy with \cite{celeghini2013-1}, an arbitrary vector 
$|f\rangle \, \in L^2(\E,\Z,\Z)$
can be expressed as
\[
|f\rangle \;=\; \sum_{m,q=-\infty}^{\infty} \int_{-1}^{+1} dx\; |x,m,q\rangle \, f^{m,q}(x) \;=
\sum_{m,q=-\infty}^{+\infty} \;\sum_{l=sup(|m|,|q|)}^\infty |l,m\rangle \;f^m_l
\]
where
\[\begin{array}{lllll}
f^{m,q}(x)&:=& \langle x,m,q|f\rangle&= &\ds \sum_{l=sup(|m|,|q|)}^\infty {\cal J}_l^{m,q}(x)\, f_l^{m,q} ,\\[0.5cm]
f_l^{m,q}& :=& \langle l,m,q|f\rangle &=&\ds \int_{-1}^{+1}dx\; {\cal J}_l^{m,q}(x)\, f^{m,q}(x)\; . 
\end{array}\]
Thus, all the $L^2$--functions defined on $(\E,\Z,\Z)$ can be written as
\[   
\sum_{m=-\infty}^{\infty}\;\sum_{q=-\infty}^{\infty}\;\sum_{l=sup(|m|,|q|)}^\infty  \;   {\cal J}_l^{m,q}(x)\; f_l^{m,q}  .
\]
Hence, they
support the UIR of $su(2,2)$.
The space of all linear operators that act on $L^2(\E,\Z,\Z)$ is thus homomorphic the UEA of $su(2,2)$.


\sect{Conclusions}\label{conclusionssection}

The relevance of the ASF seems to be related to:
\begin{enumerate}
\item
 To the role of intertwining between second order differential equations and
Lie algebras played by the algebraic special functions.

\item 
 To the fact that ASF (in this paper AJF) are at the same time an 
irreducible representation of a Lie algebra (here $su(2,2)$) and a basis of $L^2$ (and wave)
functions (here the $L^2$ functions are defined on $(\E \times \Z \times \Z/2))$.
Thus they allow to state a homorphism between the UEA constructed on the Lie algebra
and the vector space of the operators defined on the $L^2$ functions.

\item 
As the ASF are a basis of a unitary irreducible representation of the corresponding Lie group also,
all sets obtained from them applying a whatever element of the Lie group
are bases in the space of the $L^2$ functions.
\end{enumerate}


\section*{Acknowledgments}

This work was partially supported  by the Ministerio de
Educaci\'on y Ciencia  of Spain (Projects FIS2009-09002 with EU-FEDER support),  by the
Universidad de Valladolid and by
INFN-MICINN (Italy-Spain).



\end{document}